\documentclass[letter]{aa}  
\usepackage{natbib}
\usepackage{graphicx}
\begin{document}
   \title{A solar surface dynamo}

   \author{A. V{\"o}gler\thanks{Present address: 
     Sterrekundig Instituut, Utrecht University,
     Postbus 80 000, 3508 TA Utrecht, The Netherlands}          
     \and M. Sch\"ussler}

   \institute{Max-Planck-Institut f\"ur Sonnensystemforschung, 
              Max-Planck-Strasse 2, 37191 Katlenburg-Lindau, Germany\\
              \email{voegler@mps.mpg.de, msch@mps.mpg.de}
             }

   \date{\today}

  \abstract
   {Observations indicate that the `quiet' solar photosphere 
    outside active regions contains considerable amounts of magnetic
    energy and magnetic flux, with mixed polarity on small scales. 
    The origin of this flux is unclear.}
   {We test whether local dynamo action of the near-surface convection 
   (granulation) can generate a significant contribution to the observed
    magnetic flux.}
   {We have carried out MHD simulations of solar surface
  convection, including the effects of strong stratification,
  compressibility, partial ionization, radiative transfer, as well as 
  an open lower boundary.}
  { Exponential growth of a weak magnetic seed field (with vanishing net
  flux through the computational box) is found in a simulation run with
  a magnetic Reynolds number of about 2600. The magnetic energy
  approaches saturation at a level of a few percent of the total kinetic
  energy of the convective motions. Near the visible
  solar surface, the (unsigned) magnetic flux density reaches at least a
  value of about 25~G. }
   {A realistic flow topology of stratified, compressible, non-helical
  surface convection without enforced recirculation is capable of
  turbulent local dynamo action near the solar surface.}
   \maketitle
%

\section{Introduction}

Magnetic fields in the quiet solar photosphere (often referred to as
`internetwork fields') are of considerable interest in connection with
the heating of the upper solar atmosphere and as a possible example of
local fast dynamo action.  Owing to their apparently `turbulent' nature,
with mixed polarity on small spatial scales, their extended range of
field strengths, and their weak polarization signals, quiet-Sun (QS)
magnetic fields are a difficult observational target. Quantitative
results from observations based upon different methods do not yet
provide a fully consistent picture \citep[e.g.,][]{Lin:Rimmele:1999,
Sanchez:Lites:2000, Khomenko:etal:2003, Sanchez:etal:2003,
Dominguez:etal:2003, Lites:Socas-Navarro:2004, Dominguez:etal:2006a,
Dominguez:etal:2006b, Trujillo:etal:2004, Sanchez-Almeida:2003,
Sanchez-Almeida:2005}, but it appears to be established that small-scale,
mixed-polarity magnetic fields are ubiquitous in the quiet Sun and
contribute significantly to the total magnetic energy and unsigned flux
in the photosphere outside active regions.

The origin of these small-scale fields is not yet fully clarified.  It
seems plausible that rise and emergence of magnetic flux from the deep
convection zone in the form of small bipoles as well as the debris from
decaying active regions contribute to the QS magnetic fields. On the
other hand, the increasing flux replenishment rates towards smaller
scales \citep{Hagenaar:etal:2003} and the absence of a significant
variation of the QS flux between solar minimum and maximum
\citep{Trujillo:etal:2004} indicate that a considerable part of the QS
magnetic fields may be produced by a self-excited turbulent dynamo
operating in the near-surface layers of the Sun
\citep[e.g.,][]{Petrovay:Szakaly:1993}.  Since the turnover time of the
corresponding dominant convection pattern (granulation) of about 10~min
is much smaller than the solar rotation period, this flow is practically
unaffected by the Coriolis force and thus exhibits no net helicity.  The
principle of such a non-helical surface dynamo has been demonstrated in
incompressible closed-box MHD simulations of thermal convection at high
Rayleigh number \citep{Cattaneo:1999, Cattaneo:etal:2003}. The question
is whether these results carry over to the case of realistic solar
granulation. This is not at all obvious since numerical simulations of
solar surface convection show that the powerful surface cooling by
radiation and the strong stratification prevent a significant
recirculation of the downflowing plasma in the near-surface layers
\citep{Stein:Nordlund:1989}. As a consequence, small-scale magnetic flux
is rapidly pumped into the deeper layers of the convection zone and is
thus lost for further amplification by the near-surface flows
\citep{Stein:Nordlund:2003}.  In a simulation with artificially closed
boundaries, such pumping is suppressed and a strong local recirculation
maintained, so that a local dynamo is efficiently sustained.

In this paper, we present the first example of dynamo action in a
realistic simulation of solar surface convection with an open lower
boundary. Our results indicate that a realistic flow topology of
strongly stratified, compressible, and non-helical surface convection
without enforced recirculation is capable of driving a turbulent local
dynamo.


\section{Numerical model}

We have used the MURaM code \citep{Voegler:etal:2005, Voegler:2003} to
carry out local-box MHD simulations of solar surface convection with
grey radiative transfer. The computational domain covers the height
range between about 800~km below and 600~km above the average height of
the visible solar surface and has a horizontal extension between about
5~Mm and 6~Mm (see Tab.~\ref{Setup}).  

The side boundaries are periodic
in both horizontal directions. 
The boundary condition at the bottom $(z=0)$
permits free in- and outflow of matter:
the upflows are assumed to be vertical, 
$v_x=v_y=0\, , \,\partial_z v_z=0$, in the 
downflows the vertical gradients are set to zero, 
$ \partial_z v_x = \partial_z v_y = \partial_z v_z =0$.
The upper boundary is closed for the flow. 
The magnetic field is assumed to be vertical at the upper and
lower boundaries: $B_x=B_y=0\, , \,\partial_z B_z=0$. Further details
of the boundary conditions are described in \citet{Voegler:etal:2005}.
Horizontal fields which get carried towards the lower
boundary in downflows nevertheless leave the simulation domain by means
of magnetic diffusion across the boundary: at any thime, the diffusive
boundary layer at the bottom adjusts its width such that the resulting
diffusive flux matches the incoming advection flux into the boundary
layer. Since the magnetic field strength in the lower part of the box
always remains significantly below the equipartition value with the
convective flows, it is thus guaranteed that the magnetic flux carried
by the downflows leaves the box unimpeded.  In order to always ensure
that the diffusive boundary layer is well resolved numerically and to
reduce the effect of the numerical boundary on the magnetic fields in
the bulk of the simulation domain, the magnetic diffusivity is increased
in a region of 150 km thickness at the bottom boundary. The lower
boundary condition also prevents the advection of horizontal magnetic
flux {\em into} the box from below. In the real Sun, of course, magnetic
flux from the deeper layers is probably advected into the surface layers
and may influence the properties of the QS magnetic field. In our
experiment, we intentionally exclude this source of magnetic flux in
order to study local dynamo action in isolation.

\begin{table}
\caption{Simulation parameters. The estimate of the magnetic Reynolds
number is based on the rms flow velocity and a length scale of 1 Mm
(granulation scale). $\Delta z$ and $\Delta x$ are the vertical
and horizontal grid spacing, respectively. }
\label{Setup}  
\centering       
\begin{tabular}{c c c c c}   
\hline       
Run & height/width & $\Delta z$/$\Delta x$ & $\eta$ & $R_{\rm m}$ \\ 
    &  $\left[ {\rm Mm} \right]$  & $\left[ {\rm km} \right]$ 
    &  $\left[ 10^{10}{\rm cm}^2{\rm s}^{-1} \right]$  & (approx.)  \\ 
\hline\hline  
   A & $1.4$/$6.0$ & $14$/$20.8$ & $11.1$  & 300 \\
   B & $1.4$/$6.0$ & $14$/$10.4$ &  $2.5$  & 1300 \\ 
   C & $1.4$/$4.86$ & $10$/$7.5$ &  $1.25$ & 2600 \\  
\hline    
\end{tabular}
\end{table}

While we have a constant value of the magnetic diffusivity in the box
(outside the region of enhanced diffusivity near the bottom), the code
uses an artificial viscosity that varies in space and time
\citep[see][]{Voegler:etal:2005}. Consequently, we can evaluate the
magnetic Reynolds number, $R_{\rm m}$, for a simulation run, but a
quantitative estimate of the hydrodynamic Reynolds number, $Re$, and of
the magnetic Prandtl number, $Pr_{\rm m} = R_{\rm m}/Re$, is difficult
to obtain.  Since both, the explicit magnetic diffusivity and the
artificial viscosity, lead to a diffusive cutoff at the scale of the
grid resolution, $Re$ and $R_{\rm m}$ are of the same order of magnitude
and thus $Pr_{\rm m}$ is about unity at the smallest resolved spatial
scales.  For smooth flows on larger scales, the artificial viscosity is
much smaller than the magnetic diffusivity, so that our `effective'
magnetic Prandtl number on these scales is smaller than unity. Given
these conditions, our results do not provide a proper basis for
commenting on the much debated question of the dependence of turbulent
dynamo action on the value of $Pr_{\rm m}$
\citep[e.g.,][]{Boldyrev:Cattaneo:2004, Ponty:etal:2005,
Schekochihin:etal:2005,Brandenburg:Subramanian:2005}.

We have carried out three simulation runs with different magnetic
Reynolds numbers, $R_{\rm m}$.  For all runs, a weak magnetic seed field
with zero net flux was introduced into a fully developed nonmagnetic
convection pattern. The seed field was purely vertical, with polarity
variations corresponding to a checkerboard-like $4\times 4$ horizontal
planform and a constant field strength of $|B_0| = 10 \;{\rm mG}$. The
box dimensions, grid resolution, and magnetic diffusivities for the runs
are given in Tab. \ref{Setup}.

\section{Results}

\ifnum 2>1
\begin{figure}
\centering
\resizebox{\hsize}{!}{\includegraphics{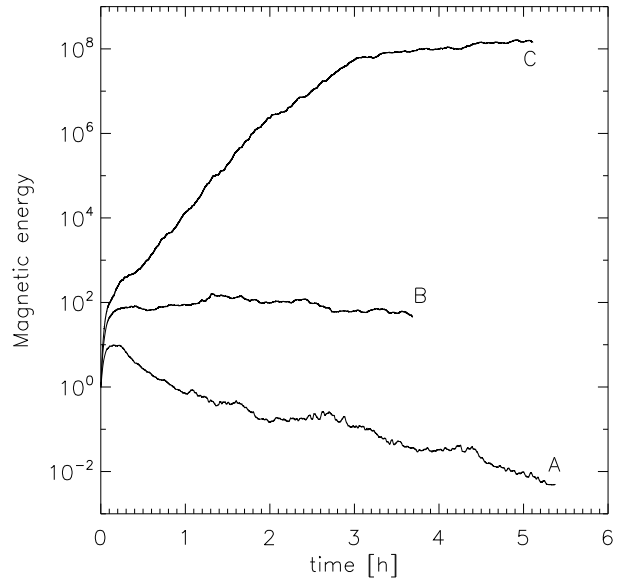}}
\caption{ Total magnetic energy (normalized to the energy of the initial
          seed field at $t=0$) in the simulation box as a function of
          simulated solar time for the three runs specified in
          Tab.~\ref{Setup}. All runs show an initial rapid energy
          increase due to flux expulsion of the seed field by
          granulation. Thereafter, run A shows an exponential decay with
          an $e$-folding time of roughly one hour, while run B is
          approximately marginal at a low energy level. Run C exhibits
          exponential growth with a time scale of about 10~minutes and
          approaches a saturation level of a few percent of the total
          kinetic energy of the convective flow.}
\label{Emag}
\end{figure}
\fi

Figure~\ref{Emag} shows the magnetic energy for the three simulation
runs as a function of time. In run A ($R_{\rm m}\approx 300$), the
magnetic energy decays exponentially after a brief initial amplification
due to flux expulsion acting on the initial magnetic configuration. Run
B ($R_{\rm m}\approx 1300$) appears to be close to the point of marginal
dynamo excitation. The magnetic energy levels out well within the
kinematic regime, with maximum field strengths several orders of
magnitude below local equipartition with respect to the kinetic energy
density of the flow. Finally, run C ($R_{\rm m} \approx 2600$) exhibits
exponential growth with an $e$-folding timescale of roughly
10~minutes. After about three hours of simulated solar time, the
magnetic energy approaches a saturated level of about 3$\%$ of the total
kinetic energy of the convective flow.

\ifnum 2>1
\begin{figure*}
\centering
\resizebox{\hsize}{!}{\includegraphics{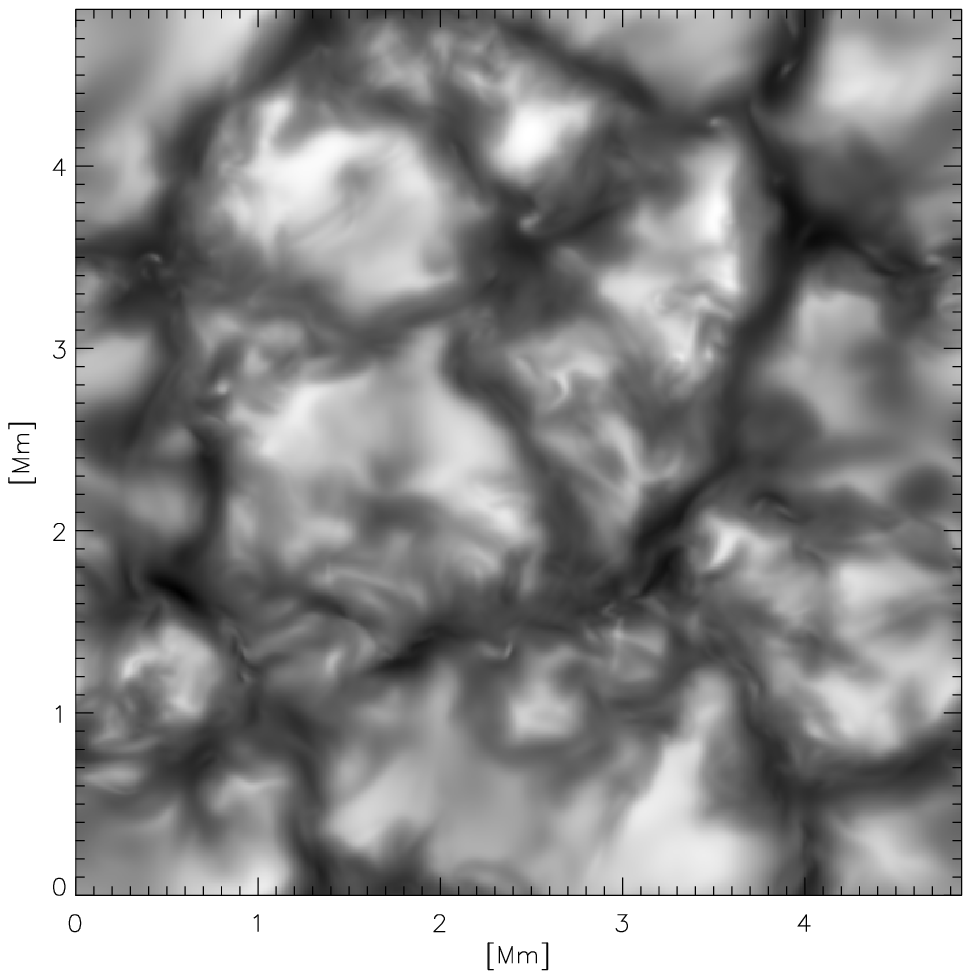}
   \includegraphics{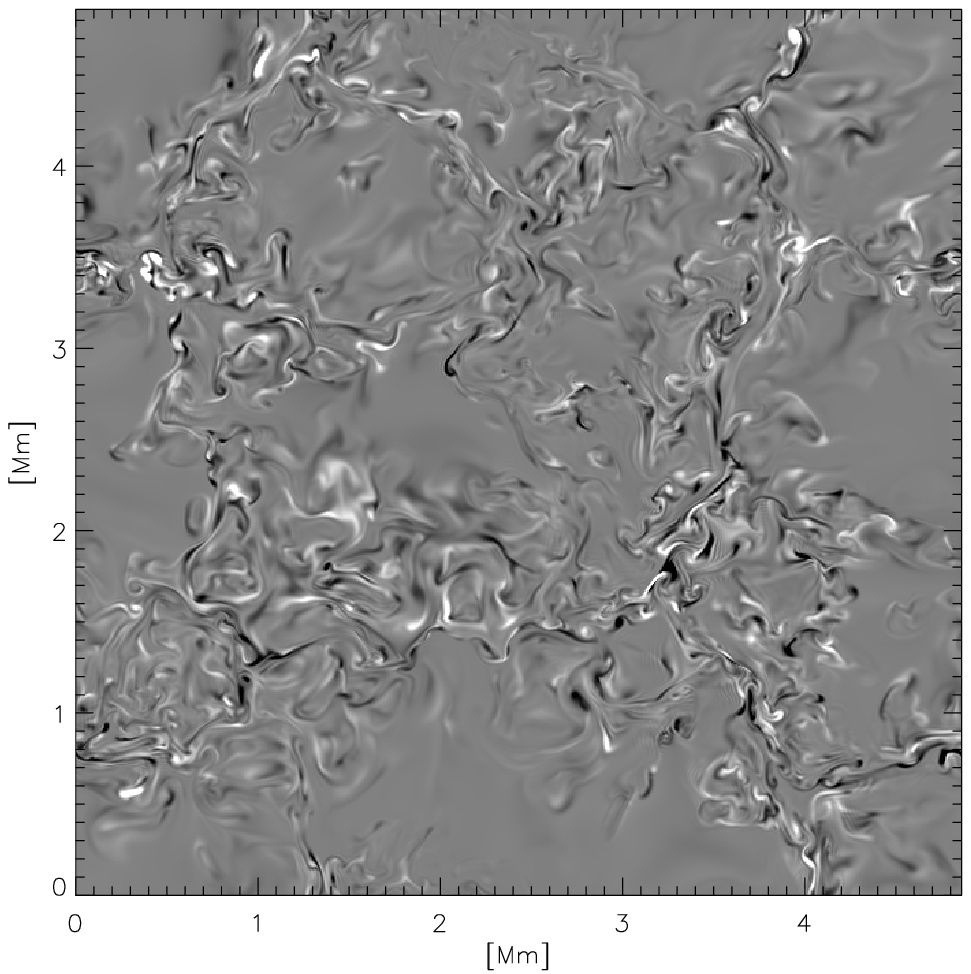}\includegraphics{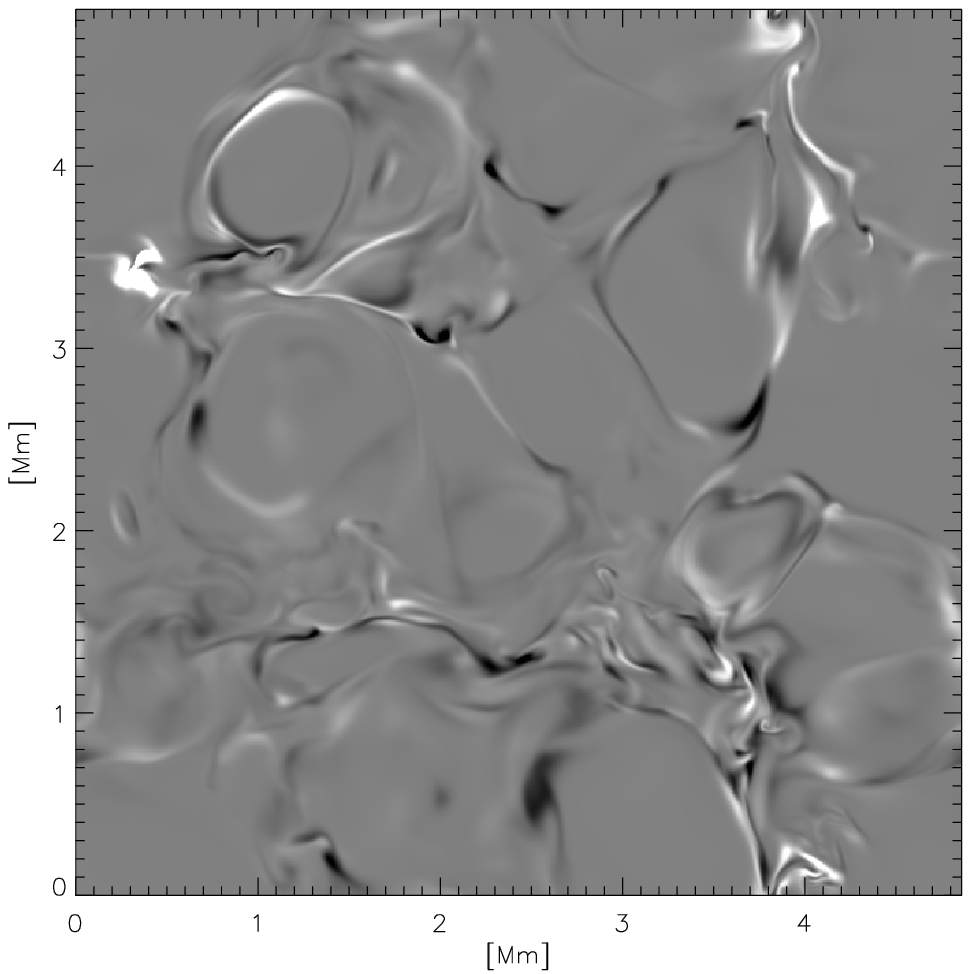}}
\caption{Snapshot from the dynamo run C, taken about 5 hours after
         introducing the seed field. The vertically emerging bolometric
         intensity (brightness, left panel) reveals a normal solar
         granulation pattern. The other panels show the vertical
         component of the magnetic field on two surfaces of constant
         (Rosseland) optical depth, $\tau_{\rm R}$. Near the visible
         surface (middle panel, $\tau_{\rm R}=1$, grey scale saturating
         at $\pm 250\,$G), the magnetic field shows an intricate
         small-scale pattern with rapid polarity changes and an unsigned
         average flux density of 25.1~G. About 300~km higher, at the
         surface $\tau_{\rm R}=0.01$ (right panel, grey scale saturating
         at $\pm 50\,$G), the unsigned average flux density has
         decreased to 3.2~G and the field distribution has become
         considerably smoother, roughly outlining the network of
         intergranular downflow lanes (darker areas on the left panel).}
\label{Maps}
\end{figure*}

\begin{figure}
\centering
\resizebox{\hsize}{!}{\includegraphics{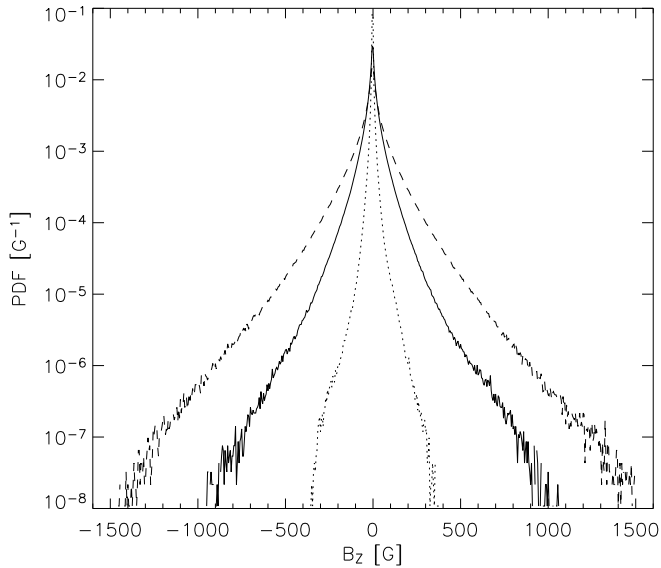}}
\caption{Probability density function (PDF) for the vertical field
component at three different geometrical height levels. Dashed curve:
$z=-370\,$km (about 450 km below the visible surface); solid curve:
$z=80\,$km (roughly corresponding to the average level of $\tau_{\rm
R}=1$); dotted curve: $z=400\,$km (about the average level of $\tau_{\rm
R}=0.01$).  Shown are time averages over about 20~minutes around
$t\simeq 4.5\,$h. }
\label{PDF}
\end{figure}
\fi

For a snapshot during the saturation phase of run C, Fig.~\ref{Maps}
shows maps of the (bolometric) brightness and of the vertical magnetic
field on two surfaces of constant (Rosseland) optical depth, $\tau_{\rm
R}$.  Around $\tau_{\rm R}=1$, the field exhibits an intricate
small-scale mixed-polarity structure, which extends down to the
diffusive length scale.  The (unsigned) mean magnetic flux density at
this level has reached a value of about 25~G. At the surface $\tau_{\rm
R}=0.01$, about 300~km higher in the atmosphere, the spatial
distribution of the field is significantly smoother and more closely
associated to the intergranular downflow lanes.  The mean flux density
has decreased to about 3~G. This strong decrease indicates the absence
of significant dynamo driving in the convectively stable layers above
$\tau_{\rm R}=1$, so that the field decays rapidly with height, owing to
its small horizontal spatial scale near the visible solar surface.

Figure~\ref{PDF} shows the average probability density function (PDF) of
the vertical magnetic field, determined during the saturation phase of
the dynamo, at three height levels.
The PDFs have the form of stretched exponentials, indicating a strong
intermittency of the magnetic field at all heights. The strongest
magnetic features occasionally reach vertical field strengths beyond
$1$~kG near $\tau_{\rm R}=1$.

Energy spectra for the vertical components of the near-surface
magnetic field and velocity as a function of horizontal wave number,
$k_{\rm h}$, are given in Fig.~\ref{spectrum}. The spectral magnetic
energy shows a broad peak at $k_{\rm h}\simeq30$, which corresponds to
a wavelength of about 200~km. At the high-wavenumber end of
the spectra, the magnetic and kinetic energies become less
disparate. The remaining deviation from equipartition is due to the
anisotropy resulting from the strong stratification.

\ifnum 2>1
\begin{figure}
\centering
\resizebox{\hsize}{!}{\includegraphics{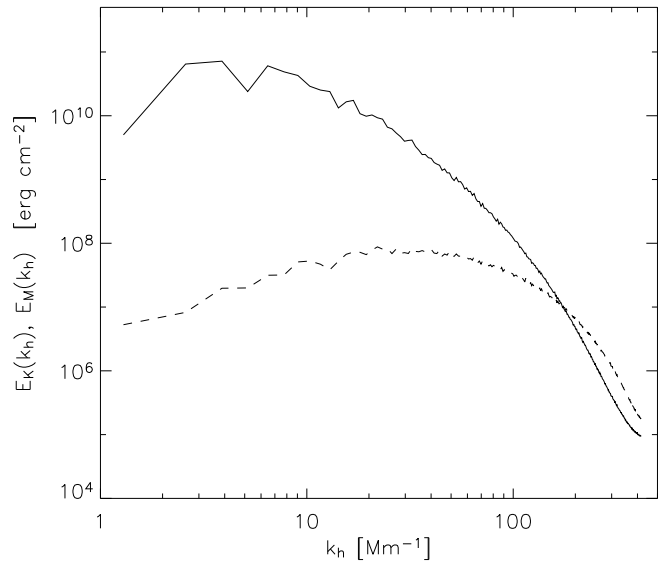}}
\caption{Energy spectra based on the vertical components of velocity
and magnetic field, respectively, as functions of horizontal wave
number, $k_{\rm h}$. The values are taken at $z=0$, corresponding to
a depth of about 80~km below the average average level of $\tau_{\rm
R}=1$), at $t\simeq 5$~h. The kinetic energy spectrum (solid curve)
peaks at $k_{\rm h}\simeq 3...4\,$Mm$^{-1}$, roughly corresponding to the
typical scale of granules. The magnetic energy spectrum has a broad
maximum around wave numbers of about $30\,$Mm$^{-1}$, corresponding to
length scales of at least an order of magnitude smaller.}
\label{spectrum}
\end{figure}
\fi

We find that convective downward pumping of flux in fact has a
significant effect on the energy balance of the dynamo, as conjectured
by \citet{Stein:Nordlund:2003}.  At any given height, the time
dependence of the horizontally averaged magnetic energy density, $e_{\rm
mag}$, is governed by the equation
$\partial_t e_{\rm mag} = W_{\rm L}-W_{\rm J} 
  - \partial\mathcal{P}_z/\partial z$. 
$W_{\rm L}$ is the rate of work against the Lorentz force, $W_{\rm J}$
is the Joule heating rate, and $\mathcal{P}_z$ is the vertical component
of the Poynting flux, the advective part of which measures the draining
of magnetic energy due to convective pumping.  All quantities are meant
to be horizontal averages.  The advective Poynting flux is negative
throughout the convectively unstable parts of the simulation domain,
confirming that any growth of magnetic energy in the system must have
its source inside the domain. The diffusive part of $\mathcal{P}_z$ is
found to be negligible in the convecting layer. In the absence of
convective pumping, the difference $W_{\rm L}-W_{\rm J}$ would be a
measure for the growth of the magnetic energy during the exponential
growth phase. In our case, more than 80\% of this difference is indeed
carried downwards by means of the term $- \partial\mathcal{P}_z/\partial
z$ and leaves the box through the bottom boundary. However, the effect
only reduces the growth rate but does not shut down dynamo action if the
magnetic Reynolds number is sufficiently large.

\section{Discussion}

Our main finding is that a realistic flow topology of strongly
stratified convection in the near-surface layers of the Sun is capable
of sustaining dynamo action. Downward pumping in an open box has a
significant impact on the energy balance, but is not able to shut down
the dynamo, presumably because there is sufficient local recirculation
(e.g. by turbulent entrainment of downflowing material into upflow
regions) to amplify the magnetic field near the surface. A more detailed
analysis will have to be carried out in order to clarify the physical
mechanism at work here.  Strictly speaking, a dynamo simulation would
have to be run for several global diffusive timescales before transients
can be definitively ruled out. This is practically not feasible in our
case since the global diffusion time for run C is about 18 days. On the
other hand, the exponential amplification of a miniscule seed field to
substantial saturation levels of flux over several hours, with a growth
time corresponding to the granulation time scale, lends credibility to
physical relevance to the results.

While demonstrating the possibility of local dynamo action by
granulation in principle, the quantitative results of our simulations
should be interpreted with caution. Owing to our subgrid model for the
viscosity, the magnetic Prandtl number cannot be uniquely defined.  We
have argued that its value is significantly smaller than unity at the
spatial scales where the magnetic energy peaks, but certainly our
numerical experiment is not suited to address the question of turbulent
dynamo action in the limit of very small $Pr_{\rm m}$.  Likewise, the
saturation level of our dynamo run probably depends on the magnetic
Reynolds number (as well as on $Pr_{\rm m}$), so that a quantitative
comparison with observational results in terms of mean flux densities
and PDFs would certainly be premature. Such comparison would also have
to take into account, in addition to the self-excited dynamo, the
advection of flux from below and the processing of magnetic debris from
decaying active regions. Still, we deem it noteworthy that levels
of magnetic flux of the observed order of magnitude can, in principle,
be produced by simulations of the kind presented here.

This is just a first step and much remains to be done in order to establish
quantitatively reliable results of near-surface local dynamo action.
This includes studying the effects of deeper and wider computational
boxes as well variations of the boundary conditions. Clearly, the dependence
of the saturation level on the magnetic Reynolds and magnetic Prandtl
numbers has to be investigated through series of controlled numerical
experiments. Comparison with observational results will be crucial in
order to evaluate the contributions by flux advection from the deeper
layers. 
     
\begin{acknowledgements}
We are grateful to Robert Cameron and Matthias Rempel for stimulating
and helpful discussions.
\end{acknowledgements}

\bibliography{7253.bbl}

\begin{thebibliography}{23}
\expandafter\ifx\csname natexlab\endcsname\relax\def\natexlab#1{#1}\fi

\bibitem[{{Boldyrev} \& {Cattaneo}(2004)}]{Boldyrev:Cattaneo:2004}
{Boldyrev}, S. \& {Cattaneo}, F. 2004, Phys. Rev. Lett., 92, 144501

\bibitem[{{Brandenburg} \& {Subramanian}(2005)}]{Brandenburg:Subramanian:2005}
{Brandenburg}, A. \& {Subramanian}, K. 2005, \physrep, 417, 1

\bibitem[{{Cattaneo}(1999)}]{Cattaneo:1999}
{Cattaneo}, F. 1999, \apj, 515, L39

\bibitem[{{Cattaneo} {et~al.}(2003){Cattaneo}, {Emonet}, \&
  {Weiss}}]{Cattaneo:etal:2003}
{Cattaneo}, F., {Emonet}, T., \& {Weiss}, N. 2003, \apj, 588, 1183

\bibitem[{{Dom{\'{\i}}nguez Cerde{\~ n}a} {et~al.}(2003){Dom{\'{\i}}nguez
  Cerde{\~ n}a}, {Kneer}, \& {S{\' a}nchez Almeida}}]{Dominguez:etal:2003}
{Dom{\'{\i}}nguez Cerde{\~ n}a}, I., {Kneer}, F., \& {S{\' a}nchez Almeida}, J.
  2003, \apj, 582, L55

\bibitem[{{Dom{\'{\i}}nguez Cerde{\~n}a}
  {et~al.}(2006{\natexlab{a}}){Dom{\'{\i}}nguez Cerde{\~n}a}, {Almeida}, \&
  {Kneer}}]{Dominguez:etal:2006a}
{Dom{\'{\i}}nguez Cerde{\~n}a}, I., {Almeida}, J.~S., \& {Kneer}, F.
  2006{\natexlab{a}}, \apj, 646, 1421

\bibitem[{{Dom{\'{\i}}nguez Cerde{\~n}a}
  {et~al.}(2006{\natexlab{b}}){Dom{\'{\i}}nguez Cerde{\~n}a}, {S{\'a}nchez
  Almeida}, \& {Kneer}}]{Dominguez:etal:2006b}
{Dom{\'{\i}}nguez Cerde{\~n}a}, I., {S{\'a}nchez Almeida}, J., \& {Kneer}, F.
  2006{\natexlab{b}}, \apj, 636, 496

\bibitem[{{Hagenaar} {et~al.}(2003){Hagenaar}, {Schrijver}, \&
  {Title}}]{Hagenaar:etal:2003}
{Hagenaar}, H.~J., {Schrijver}, C.~J., \& {Title}, A.~M. 2003, \apj, 584, 1107

\bibitem[{{Khomenko} {et~al.}(2003){Khomenko}, {Collados}, {Solanki}, {Lagg},
  \& {Trujillo Bueno}}]{Khomenko:etal:2003}
{Khomenko}, E.~V., {Collados}, M., {Solanki}, S.~K., {Lagg}, A., \& {Trujillo
  Bueno}, J. 2003, \aap, 408, 1115

\bibitem[{{Lin} \& {Rimmele}(1999)}]{Lin:Rimmele:1999}
{Lin}, H. \& {Rimmele}, T. 1999, \apj, 514, 448

\bibitem[{{Lites} \& {Socas-Navarro}(2004)}]{Lites:Socas-Navarro:2004}
{Lites}, B.~W. \& {Socas-Navarro}, H. 2004, \apj, 613, 600

\bibitem[{{Petrovay} \& {Szakaly}(1993)}]{Petrovay:Szakaly:1993}
{Petrovay}, K. \& {Szakaly}, G. 1993, \aap, 274, 543

\bibitem[{{Ponty} {et~al.}(2005){Ponty}, {Mininni}, {Montgomery}, {Pinton},
  {Politano}, \& {Pouquet}}]{Ponty:etal:2005}
{Ponty}, Y., {Mininni}, P.~D., {Montgomery}, D.~C., {et~al.} 2005, Phys. Rev.
  Lett., 94, 164502

\bibitem[{{S{\'a}nchez Almeida}(2003)}]{Sanchez-Almeida:2003}
{S{\'a}nchez Almeida}, J. 2003, \aap, 411, 615

\bibitem[{{S{\'a}nchez Almeida}(2005)}]{Sanchez-Almeida:2005}
{S{\'a}nchez Almeida}, J. 2005, \aap, 438, 727

\bibitem[{{S{\'a}nchez Almeida} {et~al.}(2003){S{\'a}nchez Almeida},
  {Dom{\'{\i}}nguez Cerde{\~n}a}, \& {Kneer}}]{Sanchez:etal:2003}
{S{\'a}nchez Almeida}, J., {Dom{\'{\i}}nguez Cerde{\~n}a}, I., \& {Kneer}, F.
  2003, \apjl, 597, L177

\bibitem[{{S{\'a}nchez Almeida} \& {Lites}(2000)}]{Sanchez:Lites:2000}
{S{\'a}nchez Almeida}, J. \& {Lites}, B.~W. 2000, \apj, 532, 1215

\bibitem[{{Schekochihin} {et~al.}(2005){Schekochihin}, {Haugen}, {Brandenburg},
  {Cowley}, {Maron}, \& {McWilliams}}]{Schekochihin:etal:2005}
{Schekochihin}, A.~A., {Haugen}, N.~E.~L., {Brandenburg}, A., {et~al.} 2005,
  \apjl, 625, L115

\bibitem[{{Stein} \& {Nordlund}(1989)}]{Stein:Nordlund:1989}
{Stein}, R.~F. \& {Nordlund}, A. 1989, \apj, 342, L95

\bibitem[{{Stein} \& {Nordlund}(2003)}]{Stein:Nordlund:2003}
{Stein}, R.~F. \& {Nordlund}, {\AA}. 2003, in Modelling of Stellar Atmospheres,
  IAU-Symp. 210, ed. N.~{Piskunov}, W.~W. {Weiss}, \& D.~F. {Gray} (San
  Francisco, California: Astronomical Society of the Pacific), 169

\bibitem[{{Trujillo Bueno} {et~al.}(2004){Trujillo Bueno}, {Shchukina}, \&
  {Asensio Ramos}}]{Trujillo:etal:2004}
{Trujillo Bueno}, J., {Shchukina}, N., \& {Asensio Ramos}, A. 2004, \nat, 430,
  326

\bibitem[{{V{\" o}gler} {et~al.}(2005){V{\" o}gler}, {Shelyag}, {Sch{\"
  u}ssler}, {Cattaneo}, {Emonet}, \& {Linde}}]{Voegler:etal:2005}
{V{\" o}gler}, A., {Shelyag}, S., {Sch{\" u}ssler}, M., {et~al.} 2005, \aap,
  429, 335

\bibitem[{{V{\"o}gler}(2003)}]{Voegler:2003}
{V{\"o}gler}, A. 2003, PhD thesis, University of G{\"o}ttingen, Germany,
  http://webdoc.sub.gwdg.de/diss/2004/voegler

\end{thebibliography}

\end{document}